\documentclass[twocolumn]{aastex63}
\hypersetup{colorlinks=true, linkcolor=cyan, anchorcolor=red, citecolor=blue, filecolor=red, urlcolor=blue, pdfauthor=author}
\usepackage{multirow}
\usepackage{lineno}
\usepackage{threeparttable}

\defcitealias{Tang2019}{Paper I}
\defcitealias{Tang2020}{Paper II}

\shorttitle{N-rich field stars}
\shortauthors{Yu {\it et~al.}}

\begin{document}


\title{Chemical Tagging N-rich Field Stars with High-resolution Spectroscopy}

\correspondingauthor{Baitian~Tang}
\email{tangbt@mail.sysu.edu.cn}

\author[0000-0003-1064-4435]{Jincheng~Yu}
\affiliation{School of Physics and Astronomy, Sun Yat-sen University, Zhuhai 519082, China}

\author[0000-0002-0066-0346]{Baitian~Tang}
\affiliation{School of Physics and Astronomy, Sun Yat-sen University, Zhuhai 519082, China}

\author[0000-0003-3526-5052]{J.~G.~Fernandez-Trincado}
\affiliation{Instituto de Astronom\'ia y Ciencias Planetarias, Universidad de Atacama, Copayapu 485, Copiap\'o, Chile}
\affiliation{Centro de Investigaci\'on en Astronom\'ia, Universidad Bernardo O Higgins, Avenida Viel 1497, Santiago, Chile}

\author[0000-0002-3900-8208]{Douglas~Geisler}
\affiliation{Departamento de Astronom\'{i}a, Casilla 160-C, Universidad de Concepci\'{o}n, Concepci\'{o}n, Chile}
\affiliation{Instituto de Investigaci\'on Multidisciplinario en Ciencia y Tecnolog\'ia, Universidad de La Serena. Avenida Ra\'ul Bitr\'an S/N, La Serena, Chile}
\affiliation{Departamento de Astronom\'ia, Facultad de Ciencias, Universidad de La Serena. Av.}

\author{Hongliang~Yan}
\affiliation{Key Lab of Optical Astronomy, National Astronomical Observatories, Chinese Academy of Sciences, 20A Datun Road, Beijing 100101, China}

\author{M.~Soto}
\affiliation{Instituto de Astronom\'ia y Ciencias Planetarias, Universidad de Atacama, Copayapu 485, Copiap\'o, Chile}

\begin{abstract}
	We measure chemical abundances for over 20 elements of 15 N-rich field stars with high resolution ($R \sim 30000$) optical spectra. We find that Na, Mg, Al, Si, and Ca abundances of our N-rich field stars are mostly consistent with those of stars from globular clusters (GCs). Seven stars are estimated to have [Al/Fe$]>0.5$, which is not found in most GC ``first generation'' stars. On the other hand, $\alpha$ element abundances (especially Ti) could show distinguishable differences between in situ stars and accreted stars. We discover that one interesting star, with consistently low [Mg/Fe], [Si/Fe], [Ca/Fe], [Ti/Fe], [Sc/Fe], [V/Fe], and [Co/Fe], show similar kinematic and [Ba/Eu] as other stars from the dissolved dwarf galaxy ``$Gaia$-Sausage-Enceladus''. The $\alpha$-element abundances and the iron-peak element abundances of the N-rich field stars with metallicities $-1.25 \le {\rm [Fe/H]} \le -0.95$ show consistent values with Milky Way field stars \footnote{We refer Milky Way field stars as Milky Way halo field stars unless specified in this paper.} rather than stars from dwarf galaxies, indicating that they were formed in situ. In addition, the neutron capture elements of N-rich field stars show that most of them could be enriched by asymptotic giant branch (AGB) stars with masses around $3 - 5\, M_{\odot}$.
\end{abstract}

\keywords{
	stars: chemically peculiar -- stars: abundances -- stars: evolution
}

\section{Introduction}
\label{sec:introduction}

Globular clusters (GCs) were traditionally considered as simple stellar population systems, i.e., all member stars originate from the same molecular cloud, share the same age and abundances. However, an increasing number of studies \citep[see e.g.][and references therein]{Gratton2012, Piotto2015, Milone2017, Tang2017, Tang2018} show that almost all GCs host two or more groups of stars with different chemical abundances, which is so-called multiple populations (MPs). Stars with enhanced N, Na, (sometimes He, Al, and Si), but depleted C, O, (sometimes Mg), are traditionally called ``second generation'' (SG) stars (or ``second population'' stars), distinct from the primordial ``first generation'' (FG) stars. SG stars and FG stars possibly differ in age by a few hundred Myr \citep[e.g.,][]{Conroy2012, Bekki2017} but have the same [Fe/H] abundance \citep[e.g.,][]{Carretta2010a}. Furthermore, SG stars have different radial distribution as FG stars in most GCs, generally being more centrally concentrated \citep[e.g][]{Lardo2011, Simioni2016}, although a few exceptions has also been reported: some GCs show that FG stars concentrated more in the inner region \citep[e.g.][]{Larsen2015, Vanderbeke2015, Lim2016}, while some GCs show similar spatial distribution between these two populations \citep[e.g.][]{Dalessandro2014, Miholics2015}.

Most scenarios trying to explain the MP phenomenon assume that chemically enriched SG stars were formed in light-element-polluted environment formed partially from the ejecta of FG stars. These stars are generally enriched in light elements, such as C, N, Na, etc, which are also known as enriched populations. The proposed astrophysical sites include very massive stars (VMSs), fast rotating massive stars (FRMSs), asymptotic giant branch (AGB) stars, etc, although none is currently able to explain all observational constraints \citep[e.g.,][]{Bastian2018}.

On one hand, stellar evolution of massive stars leads to the mass loss of GCs, especially for FG stars\citep[e.g.,][]{D_Ercole2008, Schaerer2011}. On the other hand, dynamical evolution, including internal two-body interaction and external tidal stripping, also contribute to the stellar loss of GCs \citep{Lamers2010}. FG stars are chemically similar to normal field stars, while SG stars show peculiar abundances in several light elements. Thus, identifying field stars with similar chemical pattern as GC SG stars could potentially add important observational constraints to help motivate theoretical scenarios.

With the help of large spectroscopic surveys, the search for these chemically peculiar stars is becoming more efficient. N is usually enhanced in GC SG stars, and relatively easier to measure in low-resolution spectroscopic survey, therefore, N-rich field stars found in early surveys are usually linked to stars escaped from GCs \citep[e.g.,][]{Martell2010, Martell2011, Koch2019}. Recently, high resolution spectroscopic surveys, like the Apache Point Observatory Galactic Evolution Experiment \citep[APOGEE; ][]{Majewski2017} have led to the discovery of a large group of N-rich field stars \citep[][]{Fernandez_Trincado2016, Fernandez_Trincado2017, Fernandez_Trincado2019, Martell2016, Schiavon2017}. In this respect, the unprecedented large number (more than $10^6$) of low-resolution stellar spectra available in the Large Sky Area Multi-Object Fiber Spectroscopic Telescope (LAMOST) survey \citep{Zhao2012} greatly increase the efficiency and accuracy of such a search. \citet{Tang2019} \citepalias[hereafter][]{Tang2019} and \citet{Tang2020} \citepalias[hereafter][]{Tang2020} have identified $\sim 100$ N-rich stars from a sample of metal-poor, red giant branch (RGB) field stars based on CN3839, CN4142, and CH4300 spectral indices using LAMOST DR3 (data release) and DR5. These N-rich field stars show similar kinematics as Galactic GCs, which are a mixture of GCs formed in situ and ex situ \citep[e.g.,][]{Massari2019}. To further explore the connection between N-rich field stars and GCs using the idea of ``chemical tagging'', we investigate in optical high resolution spectra to carefully examine more than 20 elements in this work. Several elements that we shown in this work (e.g., Na, Ba, Eu) are important for constraining the nature of MPs, but rarely available in near-infrared studies.

The paper is organized as follows. In Section~\ref{sec:sample}, we describe our data and reduction processes. We carefully compare our derived abundances with literature values of Milky Way (MW) GCs, MW (disc and halo stars) and dwarf galaxies to investigate the origin of N-rich field stars in Section~\ref{sec:result} and \ref{sec:discussion}. Our final conclusion is given in Section~\ref{sec:conclusion}.

\section{The Data}
\label{sec:sample}

\subsection{Follow-up observations and data reduction}

We started the selection of observable targets from our LAMOST N-rich field star sample (\citetalias[][]{Tang2019} and \citetalias[][]{Tang2020}), which are mostly located in the northern sky. Fifteen stars with DEC $<10^{\circ}$ and a reasonable airmass were selected. Our follow-up observations were taken using the MIKE \citep[Magellan Inamori Kyocera Echelle,][]{Bernstein2003} spectrograph on the Magellan Clay telescope \citep{Shectman2003} at Las Campanas Observatory, Chile. The data were taken in two runs, 2019/07 and 2020/02. The blue and red detectors were used simultaneously to cover $3200<\lambda<5000$ \AA~(blue side) and $4900<\lambda<10000$ \AA~(red side). We used the 0.7'' slit and 2x2 spatial on-chip binning to achieve nominal spectral resolution of 35k/28k on the blue side and red side, respectively. Depending on the observation schedule and the brightness of the star, we took several exposures to ensure SNR (signal-to-noise ratio) $>50$ over most wavelength of the red detector (Table~\ref{tab:stellar_parameters}).

We reduced the observational data using the {\emph {CarPy}} package \citep{Kelson2000, Kelson2003}. The observed spectral images were bias-subtracted, flat-field corrected, wavelength calibrated, scatter-light and sky subtracted. The multi-order reduced spectra were then merged into a single spectrum.

\subsection{Stellar parameters and chemical abundances}

Stellar parameters and chemical abundances were analyzed using Brussels Automatic Stellar Parameter (BACCHUS) code \citep{Masseron2016}. We briefly describe the basic procedures here, readers are referred to \citet{Masseron2016} for detailed descriptions.

The equivalent widths (EW) of Fe I and Fe II absorption lines are first measured. To derive stellar parameters, several stellar parameters are attempted before they converge: $T_{\rm eff}$ is determined by obtaining null trend for the abundance of iron lines against excitation potential (see Figure~\ref{fig:Teff_determ} in Appendix); $\log g$ is obtained by ionisation equilibrium between Fe I and Fe II lines; metallicity is obtained by the mean of individual line abundances relative to the solar value; microturbulence velocity is determined by obtaining null trend for the abundance of iron lines against equivalent widths. For initial guess, we used photometric $T_{\rm eff}$ calculated from 2MASS photometry \citep{GH2009} and $\log g$, [Fe/H] from LAMOST suggested values. Our derived stellar parameters using BACCHUS are shown in Table~\ref{tab:stellar_parameters}. Note that most of our stars show $\log g$ greater than 2. Because the red clump feature is located at $\log g \sim 2$ \citepalias{Tang2020}, our sample stars are mostly RGB stars.

\begin{table*}
	\caption{Stellar parameters}
	\label{tab:stellar_parameters}
	\begin{center}
		\scriptsize
\begin{tabular}{c c c c c c c c c c c}
\hline
\hline
\hline
	\multirow{2}{*}{Id} & \multirow{2}{*}{RA} & \multirow{2}{*}{DEC} &
	$T_{\rm eff}$ & $\log g$ & [Fe/H] & $G_{\rm mag}$ &
	microturbulence        & $RV$                   &
	\multirow{2}{*}{airmass} & exposure time \\
	                    &                     &                      &
	(K)           & (dex)    & (dex)  & (mag)         &
	${\rm \,(km\,s^{-1})}$ & ${\rm \,(km\,s^{-1})}$ &
	                    & (s)                \\
\hline
\hline
	 9 (II) &  58.221019 &  7.203154 & 4312 & 0.65 & -1.34 & 13.63 & 1.78 & 69.0   & 1.42 & 4800 \\
	38 (II) & 258.215699 &  8.130489 & 4604 & 0.73 & -1.62 & 12.57 & 2.04 & -146.0 & 1.90 & 3600 \\
	47 (II) & 214.054033 & -2.145359 & 4953 & 2.14 & -1.38 & 12.06 & 1.95 & 66.5   & 1.80 & 3600 \\
	58 (II) & 197.656039 & -6.979531 & 5213 & 2.84 & -1.45 & 14.45 & 1.78 & -27.0  & 1.10 & 5400 \\
	62 (II) & 122.120480 &  1.946907 & 5020 & 2.36 & -1.42 & 12.87 & 1.56 & 147.0  & 1.17 & 5400 \\
	65 (II) & 233.551930 &  0.326274 & 4806 & 1.69 & -1.63 & 12.76 & 1.45 & -72.5  & 1.57 & 3600 \\
	67 (I)  & 317.852325 & -2.385546 & 5265 & 2.12 & -0.97 & 13.06 & 2.46 & -5.0   & 1.14 & 3600 \\
	69 (I)  & 150.981537 &  1.298311 & 5129 & 3.02 & -1.07 & 12.22 & 1.26 & 50.0   & 1.12 & 3600 \\
	80 (I)  & 134.988876 &  1.055260 & 4800 & 1.37 & -1.88 & 13.79 & 1.35 & 179.0  & 1.00 & 5400 \\
	82 (I)  & 195.094070 & -7.636771 & 5017 & 1.90 & -1.06 & 12.59 & 1.60 & -6.0   & 1.50 & 5400 \\
	86 (I)  & 239.791901 & 10.020472 & 5149 & 2.91 & -0.98 & 14.10 & 1.37 & -141.0 & 1.47 & 3600 \\
	88 (I)  & 244.844238 &  9.523706 & 4879 & 2.31 & -1.02 & 13.16 & 1.51 & -127.0 & 1.88 & 3600 \\
	94 (I)  & 166.054535 &  8.444023 & 5161 & 2.88 & -1.18 & 14.44 & 1.33 & 51.0   & 2.00 & 3600 \\
	97 (I)  & 222.293640 & 10.045577 & 4762 & 2.20 & -1.25 & 13.12 & 1.34 & -31.0  & 1.49 & 3600 \\
	98 (I)  & 199.145416 & 10.957638 & 4529 & 0.77 & -1.07 & 10.59 & 2.03 & 84.0   & 1.30 & 3600 \\
\hline
\hline
\hline
\end{tabular}

	\end{center}
	\raggedright{\hspace{0cm} The {\it Id} is the corresponding id in \citetalias{Tang2020} (table 3), with (I) and (II) indicate that they are from \citetalias{Tang2019} and \citetalias{Tang2020}, respectively.}\\
\end{table*}

To derive chemical abundance for a given absorption line, a sigma-clipping is first applied on the selected continuum points around the targeted line, then a linear fit is used for the remaining points as the continuum. Therefore, the code can detect significantly bad fits, like a sudden drop in the spectrum due to bad pixels in the detectors. Observed spectra and model spectra are compared in four different methods to determine abundances: $\chi^2$ minimization, line intensity, equivalent width, and spectral synthesis (see Figure~\ref{fig:spec_abund} in Appendix). Furthermore, the code gives each of them a flag to indicate the estimation quality. When multiple transition lines are detected for a given element, the lines that satisfy the followings are chosen: 1) $\lambda>5000\,\text{\AA}$ (for better SNR, see Figure~\ref{fig:spectra} in Appendix); 2) all four determination methods are flagged as good ($=1$); 3) SNR$>40$. Then, the mean value of the chemical abundances of different absorption lines from $\chi^2$ minimization is given as the estimated abundance for the given element. If no line meets the above criteria, the abundance is not estimated. The derived chemical abundances of our targets are given in Table~\ref{tab:abundances}.

\begin{table*}
	\caption{Chemical abundances}
	\label{tab:abundances}
	\begin{center}
		\scriptsize
\begin{tabular}{c c c c c c c c c}
\hline
\hline
\hline
	Element & 9 (II) & 38 (II) & 47 (II) & 58 (II) & 62 (II) & 65 (II) & 67 (I) & 69 (I) \\
\hline
\hline
	${\rm [O/Fe] }$ & $< 0.14$ & $<-0.02$ &    0.57  & $< 0.41$ & $<-0.18$ & $< 0.16$ &    0.49  & $< 0.37$ \\
	${\rm [Na/Fe]}$ &    0.14  &    0.38  &    0.33  &    0.52  &    0.01  &  \nodata &    0.16  &    0.16  \\
	${\rm [Mg/Fe]}$ &    0.19  &    0.31  &    0.45  &    0.31  &   -0.38  &    0.31  &    0.41  &    0.41  \\
	${\rm [Al/Fe]}$ &    0.21  &    1.29  &    0.28  &    0.39  &    0.68  &    0.83  &    0.26  &  \nodata \\
	${\rm [Si/Fe]}$ &    0.11  &    0.51  &    0.14  &    0.30  &    0.45  &    0.45  &    0.41  &    0.35  \\
	${\rm [S/Fe] }$ &  \nodata &    0.55  &  \nodata &    0.59  &  \nodata &  \nodata &    0.20  &    0.52  \\
	${\rm [Ca/Fe]}$ &    0.11  &    0.44  &    0.27  &    0.33  &    0.38  &    0.39  &    0.10  &    0.41  \\
	${\rm [Sc/Fe]}$ &   -0.23  &    0.09  &    0.19  &    0.15  &    0.34  &   -0.04  &   -0.17  &    0.41  \\
	${\rm [Ti/Fe]}$ &   -0.02  &    0.21  &    0.41  &    0.31  &    0.17  &    0.35  &    0.22  &    0.34  \\
	${\rm [V/Fe] }$ &   -0.22  &    0.05  &    0.22  &    0.23  &   -0.09  &    0.10  &    0.10  &    0.03  \\
	${\rm [Cr/Fe]}$ &   -0.12  &   -0.15  &   -0.06  &   -0.14  &   -0.24  &  \nodata &   -0.28  &   -0.03  \\
	${\rm [Mn/Fe]}$ &   -0.60  &   -0.36  &   -0.37  &   -0.39  &   -0.60  &   -0.63  &   -0.36  &   -0.50  \\
	${\rm [Co/Fe]}$ &   -0.26  &   -0.02  &    0.08  &    0.10  &   -0.01  &  \nodata &    0.13  &   -0.05  \\
	${\rm [Ni/Fe]}$ &   -0.29  &   -0.02  &    0.02  &    0.04  &   -0.14  &   -0.07  &    0.13  &   -0.01  \\
	${\rm [Y/Fe] }$ &   -0.08  &   -0.09  &   -0.05  &   -0.04  &   -0.01  &    0.04  &   -0.10  &    0.22  \\
	${\rm [Zr/Fe]}$ &   -0.00  &    0.22  &    0.42  &  \nodata &    0.35  &  \nodata &    0.31  &  \nodata \\
	${\rm [Ba/Fe]}$ &   -0.17  &   -0.20  &   -0.02  &    0.22  &    0.23  &    0.89  &    0.28  &    0.38  \\
	${\rm [La/Fe]}$ &    0.07  &    0.08  &    0.30  &    0.33  &    0.31  &    0.71  &    0.26  &    0.38  \\
	${\rm [Ce/Fe]}$ &   -0.09  &   -0.05  &    0.15  &    0.35  &    0.24  &  \nodata &    0.12  &    0.20  \\
	${\rm [Nd/Fe]}$ &    0.18  &    0.18  &    0.39  &    0.76  &    0.37  &  \nodata &    0.28  &  \nodata \\
	${\rm [Eu/Fe]}$ &    0.41  &    0.29  &    0.65  &    0.55  &    0.53  &  \nodata &    0.28  &    0.27  \\
\hline
\hline
	Element & 80 (I) & 82 (I) & 86 (I) & 88 (I) & 94 (I) & 97 (I) & 98 (I) \\
\hline
\hline
	${\rm [O/Fe] }$ &    0.26  &    0.46  &    0.95  &    0.56  & $< 0.39$ & $< 0.63$ &    0.61  \\
	${\rm [Na/Fe]}$ &    0.96  &    0.34  &    0.22  &    0.23  &    0.22  &    0.30  &    0.21  \\
	${\rm [Mg/Fe]}$ &    0.72  &    0.51  &    0.19  &    0.27  &    0.54  &    0.24  &    0.48  \\
	${\rm [Al/Fe]}$ &    0.74  &    0.20  &    0.21  &    0.14  &    0.74  &    1.01  &  \nodata \\
	${\rm [Si/Fe]}$ &    0.48  &    0.55  &    0.16  &    0.26  &    0.42  &    0.54  &    0.43  \\
	${\rm [S/Fe] }$ &    0.82  &    0.27  &  \nodata &  \nodata &    0.59  &  \nodata &  \nodata \\
	${\rm [Ca/Fe]}$ &    0.63  &    0.41  &    0.32  &    0.23  &    0.54  &    0.60  &    0.16  \\
	${\rm [Sc/Fe]}$ &    0.10  &    0.08  &    0.30  &    0.20  &  \nodata &    0.14  &   -0.39  \\
	${\rm [Ti/Fe]}$ &    0.36  &    0.25  &    0.36  &    0.22  &    0.43  &    0.50  &    0.04  \\
	${\rm [V/Fe] }$ &    0.16  &    0.07  &    0.11  &   -0.05  &    0.20  &    0.19  &   -0.22  \\
	${\rm [Cr/Fe]}$ &   -0.02  &   -0.03  &    0.08  &   -0.16  &    0.10  &    0.18  &   -0.15  \\
	${\rm [Mn/Fe]}$ &   -0.39  &   -0.37  &   -0.38  &   -0.51  &   -0.32  &   -0.32  &   -0.49  \\
	${\rm [Co/Fe]}$ &    0.04  &    0.01  &    0.10  &   -0.17  &    0.05  &    0.06  &   -0.01  \\
	${\rm [Ni/Fe]}$ &    0.15  &    0.03  &    0.04  &   -0.09  &    0.08  &    0.21  &   -0.01  \\
	${\rm [Y/Fe] }$ &   -0.06  &    0.16  &    0.16  &    0.08  &    0.17  &    0.33  &  \nodata \\
	${\rm [Zr/Fe]}$ &    0.34  &    0.43  &  \nodata &  \nodata &  \nodata &    0.67  &  \nodata \\
	${\rm [Ba/Fe]}$ &    0.31  &  \nodata &    0.44  &    0.46  &    0.40  &    0.76  &  \nodata \\
	${\rm [La/Fe]}$ &    0.21  &    0.45  &    0.74  &    0.63  &    0.42  &    0.46  &   -0.12  \\
	${\rm [Ce/Fe]}$ &    0.16  &    0.33  &    0.29  &    0.32  &    0.22  &  \nodata &   -0.14  \\
	${\rm [Nd/Fe]}$ &    0.29  &    0.35  &    0.82  &    0.49  &    0.51  &    0.48  &    0.02  \\
	${\rm [Eu/Fe]}$ &  \nodata &    0.50  &    0.90  &  \nodata &    0.38  &  \nodata &  \nodata \\
\hline
\hline
\hline
\end{tabular}

	\end{center}
\end{table*}

In this work, we analyzed the internal errors by propagating the typical errors in $T_{\rm eff}$, $\log g$, ${\rm [Fe/H]}$. The typical errors are set as $\Delta T_{\rm eff} = 50\,{\rm K}$, $\Delta \log g = 0.1\,{\rm dex}$, $\Delta {\rm [Fe/H]} = 0.05\,{\rm dex}$. The total estimated error is thus calculated as: $\sigma_{\rm tot} = \left( (\sigma_{\Delta T_{\rm eff}})^2 + (\sigma_{\Delta \log g})^2 + (\sigma_{\Delta {\rm [Fe/H]}})^2 \right)^{1/2}$. The abundance errors of a typical star (star \#88 from \citetalias{Tang2020}) are shown in Table~\ref{tab:abu_err} as an example.

\begin{table*}
	\caption{Errors of chemical abundances propagated from atmospheric parameters for star \#88.}
	\label{tab:abu_err}
	\begin{center}
		\newcommand{\tabincell}[2]{\begin{tabular}{@{}#1@{}}#2\end{tabular}}
\scriptsize
\begin{tabular}{c c c c c c}
\hline
\hline
\hline
	Element & abundance & $\Delta T_{\rm eff}=50 {\rm \,(K)}$ & $\Delta \log (g) = 0.1$\,(dex) & $\Delta {\rm [Fe/H]} = 0.05$\,(dex) & $\sigma{\rm tot}$ \\
\hline
\hline
	$\Delta({\rm [O/Fe] })$ &   0.56  &   0.00  &   0.03  &   0.03  &   0.04  \\
	$\Delta({\rm [Na/Fe]})$ &   0.23  &   0.03  &   0.00  &   0.08  &   0.09  \\
	$\Delta({\rm [Mg/Fe]})$ &   0.27  &   0.03  &   0.00  &   0.00  &   0.03  \\
	$\Delta({\rm [Al/Fe]})$ &   0.14  &   0.02  &   0.00  &   0.02  &   0.03  \\
	$\Delta({\rm [Si/Fe]})$ &   0.26  &   0.01  &   0.01  &   0.02  &   0.03  \\
	$\Delta({\rm [S/Fe] })$ & \nodata & \nodata & \nodata & \nodata & \nodata \\
	$\Delta({\rm [Ca/Fe]})$ &   0.23  &   0.04  &   0.00  &   0.00  &   0.04  \\
	$\Delta({\rm [Sc/Fe]})$ &   0.20  &   0.00  &   0.03  &   0.02  &   0.04  \\
	$\Delta({\rm [Ti/Fe]})$ &   0.22  &   0.04  &   0.01  &   0.01  &   0.04  \\
	$\Delta({\rm [V/Fe] })$ &  -0.05  &   0.07  &   0.00  &   0.00  &   0.07  \\
	$\Delta({\rm [Cr/Fe]})$ &  -0.16  &   0.03  &   0.02  &   0.04  &   0.06  \\
	$\Delta({\rm [Mn/Fe]})$ &  -0.51  &   0.08  &   0.01  &   0.01  &   0.08  \\
	$\Delta({\rm [Co/Fe]})$ &  -0.17  &   0.06  &   0.01  &   0.01  &   0.07  \\
	$\Delta({\rm [Ni/Fe]})$ &  -0.09  &   0.03  &   0.01  &   0.03  &   0.04  \\
	$\Delta({\rm [Y/Fe] })$ &   0.08  &   0.04  &   0.04  &   0.05  &   0.07  \\
	$\Delta({\rm [Zr/Fe]})$ & \nodata & \nodata & \nodata & \nodata & \nodata \\
	$\Delta({\rm [Ba/Fe]})$ &   0.46  &   0.01  &   0.05  &   0.06  &   0.08  \\
	$\Delta({\rm [La/Fe]})$ &   0.63  &   0.01  &   0.04  &   0.01  &   0.05  \\
	$\Delta({\rm [Ce/Fe]})$ &   0.32  &   0.01  &   0.02  &   0.03  &   0.04  \\
	$\Delta({\rm [Nd/Fe]})$ &   0.49  &   0.03  &   0.04  &   0.07  &   0.09  \\
	$\Delta({\rm [Eu/Fe]})$ & \nodata & \nodata & \nodata & \nodata & \nodata \\
\hline
\hline
\hline
\end{tabular}

	\end{center}
\end{table*}

\section{Results}
\label{sec:result}

\subsection{Na-O relation}
\label{subsec:NaO}

The anti-correlation between Na and O is observed in almost all GCs \citep[][]{Carretta2009a, Carretta2010a, Gratton2012}, which is arguably the most characteristic MP property of GC member stars. The Na-O anti-correlation is suggested to originate from the CNO cycle and the NeNa cycle activated during lower temperature H-burning. If our N-rich field stars show compatible Na and O abundances as the GC enriched populations, it would be a supportive evidence for their GC origin scenario. For the reference of chemical abundances of GC stars, we used the observed data of 19 GCs from \citet{Carretta2009a}. Based on the Na and O abundances, they separated members of each GC into three components, P (primordial) component, I (intermediate) component, and E (extreme) component. FG stars (or P component) are assumed to have similar O and Na abundances as field stars of the same metallicity, while I and E components are all considered to be SG stars.

\begin{figure}
	\centering \includegraphics[width=1.0\linewidth,angle=0]{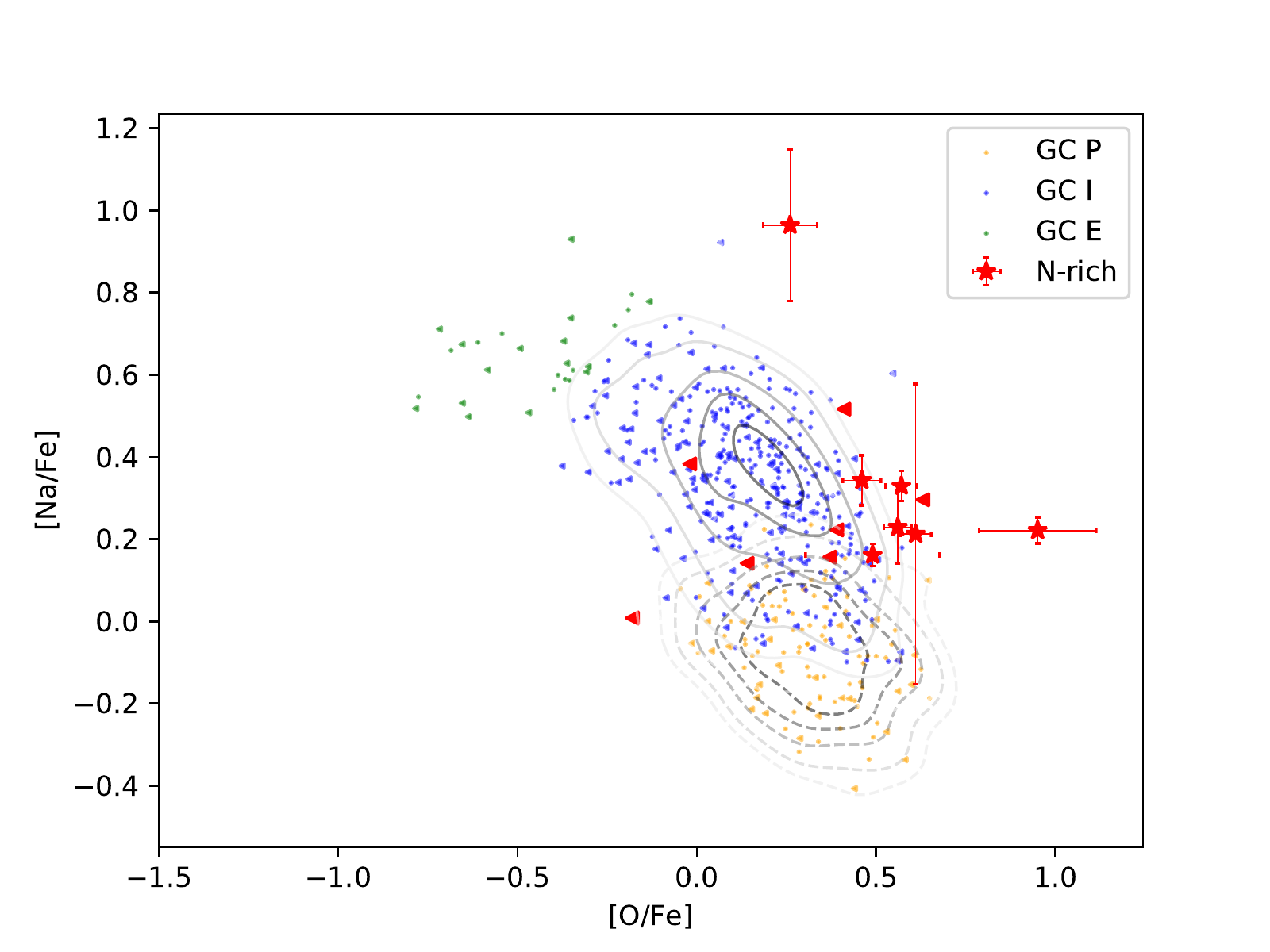}
	\caption{[Na/Fe]-[O/Fe] relation. Na and O abundances of N-rich field stars are shown as red symbols. Upper limits in O abundances are shown as triangles, while detections are shown as stars with error bars. The Na and O abundances of  the individual stars in the GCs from \citet{Carretta2009a} are over plotted for comparison. Green symbols indicate the GC E component. Blue symbols indicate the GC I component, while orange symbols indicate the GC P component. Upper limits in O abundances are shown as triangles, while detections are shown as dots. The solid and dashed contours indicate the number densities of I component and P component, respectively.}
	\label{fig:NaO}
\end{figure}

Following the convention to measure O abundances \citep[e.g.][]{Carretta2009a}, we only used [O I] $\lambda \lambda$6300, 6363$\text{\AA}$ forbidden lines to minimize the effect of non-local thermodynamic equilibrium (NLTE) correction. If no measurable line is found, the mean value from $\chi^2$ minimization is given as the upper limit. The Na and O abundances are shown in Figure~\ref{fig:NaO}, where GC stars from \citet{Carretta2009a} are shown as background dots and contours. In order to match the metallicities between our N-rich field stars and GCs, only GCs with $-1.8 \le {\rm [Fe/H]} \le -1.0$ are used. Though a clear anti-correlation between Na and O with large spread is found for GC stars, the separation between FG and SG stars is not clear-cut. Seven stars have measurable O abundances.  Most N-rich field stars basically show consistent Na and O abundances with those in GCs, except for two stars: one with [O/Fe$] \sim 1$ and the other with [Na/Fe$] \sim 1$. Based on their measurable Na abundances, most N-rich field stars are located in the transition region between FG and SG stars, thus it is difficult to verify their SG origin.

\subsection{Mg-Al relation}

\begin{figure}
	\centering \includegraphics[width=1.0\linewidth,angle=0]{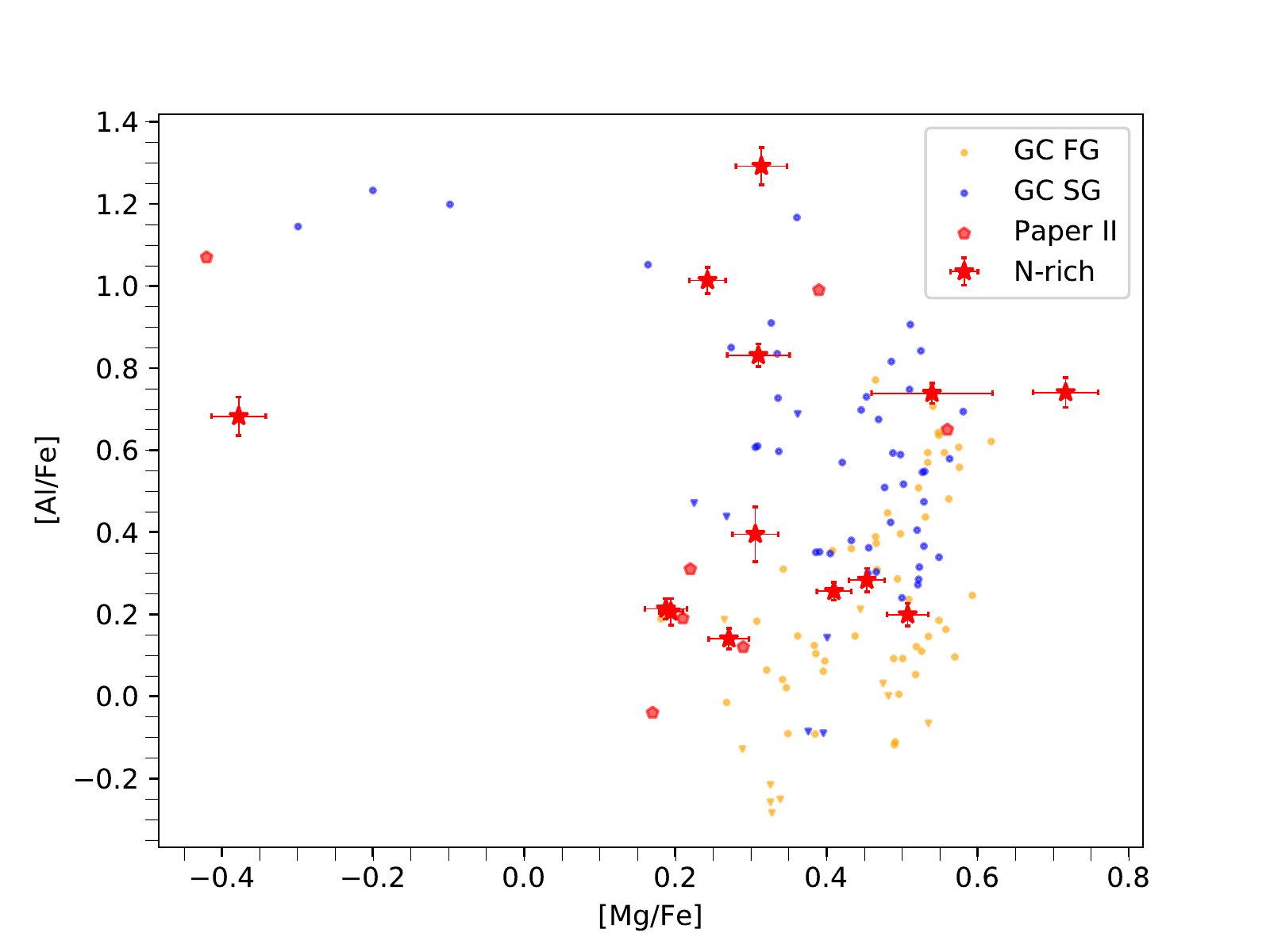}
	\caption{[Mg/Fe]-[Al/Fe] relation. Mg and Al abundances of N-rich field stars are shown as red symbols: 15 N-rich stars observed with MIKE spectra are marked as red stars with error bars, while seven N-rich stars with APOGEE spectra \citepalias{Tang2020} are marked as red pentagons. Orange and blue symbols correspond to GC first generation (P component) and second generation (I and E component) stars, respectively \citep{Carretta2009b}. Upper limits in Al abundances of GC stars are shown as triangles, while detections are shown as small dots.}
	\label{fig:MgAl}
\end{figure}

Mg-Al anti-correlation is another common feature found in low metallicity GCs (${\rm [Fe/H]} < -1.0$, \citealt{Pancino2017}), which is suggested to be the consequence of Mg-Al cycle activated at higher core temperature during H-burning \citep{Arnould1999}. Our N-rich field stars are shown as red symbols in Figure~\ref{fig:MgAl}. We note that the red pentagons are chemical abundances derived in \citetalias{Tang2020} using APOGEE spectra. In the background, GC stars from \citet{Carretta2009b} are separated in to FG (orange) and SG (blue) stars as previously mentioned.

\begin{figure}
	\centering \includegraphics[width=1.0\linewidth,angle=0]{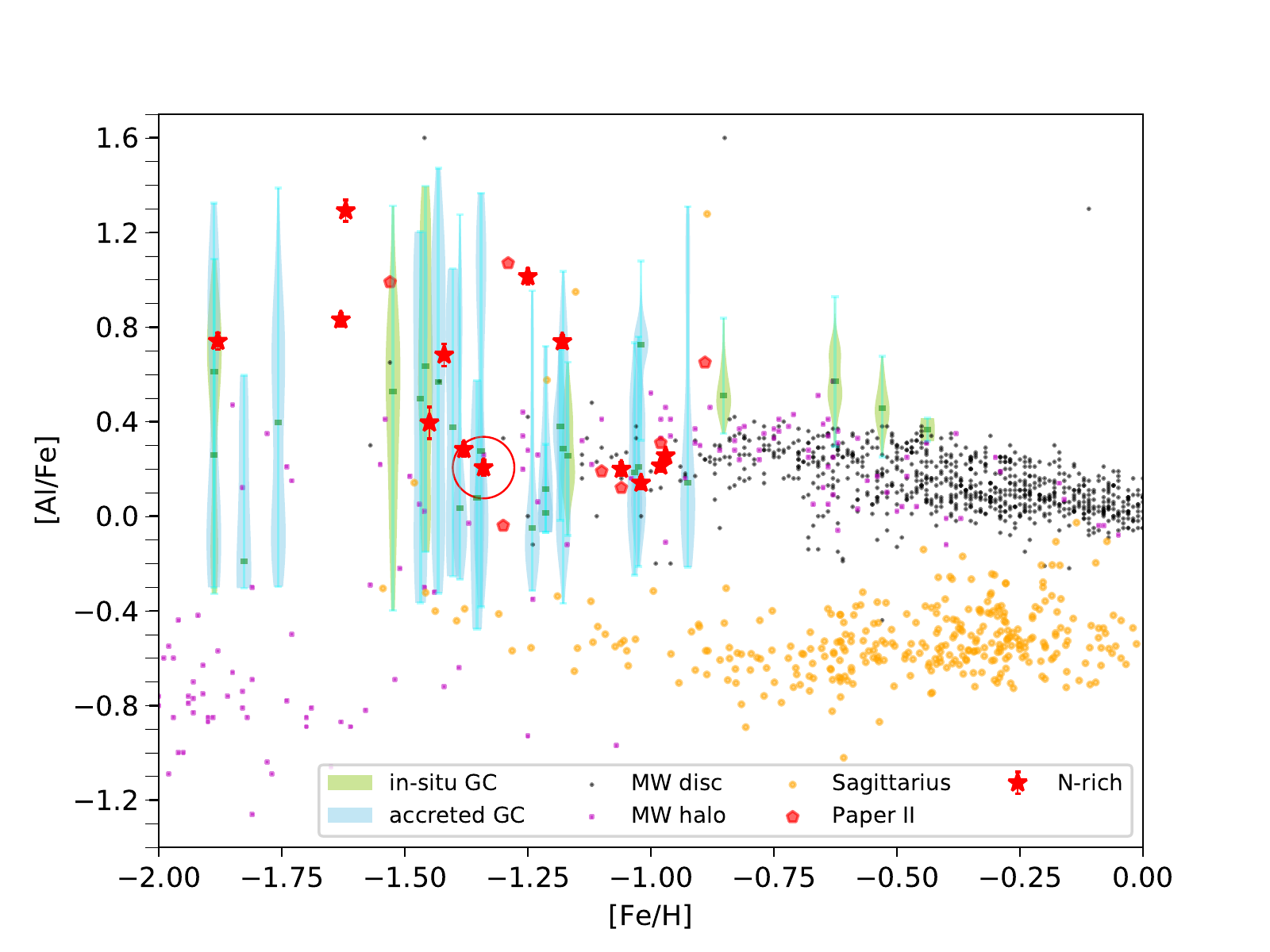}
	\caption{[Al/Fe]-[Fe/H] relation. Al abundances of N-rich sample with MIKE spectra are shown as red symbols (same meanings as mentioned in Figure~\ref{fig:MgAl}). Magenta squares correspond to MW stars from halo (92 out of 168 stars from \citealt{Fulbright2000}, 35 stars from \citealt{Cayrel2004}, 234 out of 253 stars from \citealt{Barklem2005}, 131 out of 199 stars from \citealt{Yong2013}, 287 out of 313 stars from \citealt{Roederer2014}), while black dots for MW stars from disc (174 out of 181 stars from \citealt{Reddy2003}, 153 out of 176 stars from \citealt{Reddy2006}, 679 out of 714 stars from \citealt{Bensby2014}). Orange circles correspond to Sagittarius \citep{Hasselquist2017, Hayes2020}. Violin shaped symbols indicate GC stars from APOGEE \citep{Meszaros2020}. In situ GCs are colored with yellow green, while accreted GCs are colored with sky blue. Star \#9 with high extragalactic GC origin possibility are labeled with large red circle.}
	\label{fig:Al_FeH}
\end{figure}

Though GC FG and SG stars are wide-spread and even overlap in [Al/Fe] distribution, FG stars seldomly exceed ${\rm [Al/Fe]} \sim 0.5$ (excluding upper limits). On the other hand, GC SG stars may show low [Al/Fe] down to 0, and thus a lower [Al/Fe] cannot exclude the possibility that the star belongs to SG. In this sense, almost half of our N-rich field star sample show ${\rm [Al/Fe]} > 0.5$, which are probable SG stars. Those stars with ${\rm [Al/Fe]} > 0.5$ also have compatible [Mg/Fe] to ensure that they are covered by the Mg-Al anti-correlation of GC Al-enhanced stars. There are two extremely Mg depleted N-rich field star (${\rm [Mg/Fe]} < -0.4$), but one of them has been reported in \citetalias{Tang2020} and \citet{Fernandez_Trincado2019}. The Mg depleted N-rich field stars were also discussed in \citet{Fernandez_Trincado2016, Fernandez_Trincado2017}. The Mg depletion in N-rich field stars is rare, it would be interesting to know if these Mg-depleted N-rich field stars have gone through additional nucleosynthesis. Meanwhile, the other half of N-rich field stars (${\rm [Al/Fe]} < 0.5$) show Mg and Al abundances consistent with GC FG/SG stars. Though chemically being less distinguishable from FG stars based on their Mg and Al abundances, we cannot rule out the possibilities that lower [Al/Fe] N-rich field stars are SG stars, as more metal-rich GCs tend to show similar Al abundances for FG and SG stars \citep{Pancino2017}. Furthermore, using APOGEE GC stars \citep{Meszaros2020}, we found that lower Al abundances (${\rm [Al/Fe]} < 0.5$) N-rich field stars tend to be more metal-rich in the sample (Figure~\ref{fig:Al_FeH}), tentatively agree with the aforementioned GC behavior.

\subsection{\texorpdfstring{$\alpha$}{alpha}-elements}
\label{sec:alpha}

The $\alpha$-elements are mainly generated in massive stars through type II supernovae (SNe II), consequently their recycle timescale in the interstellar medium is much smaller than that of iron, which is mainly produced in type Ia supernovae (SNe Ia). As a result, the $[\alpha/{\rm Fe}]$ abundance is initially enhanced, and starts to decline with [Fe/H] after SN Ia explosion rate reaches a maximum \citep{Matteucci1986}. The decline point, or the knee in $[\alpha/{\rm Fe}]$ vs. ${\rm [Fe/H]}$ trend, is related to star formation rate and thus galaxy mass: the less massive the galaxy is, the more metal-poor is the $[\alpha/{\rm Fe}]$ turnover. O and Mg are commonly classified as hydrostatic $\alpha$-elements, while Si, Ca, and Ti are classified as explosive $\alpha$-elements \citep{Woosley1995}, implying different nucleosynthetic processes.

\begin{figure*}
	\centering \includegraphics[width=1.0\linewidth,angle=0]{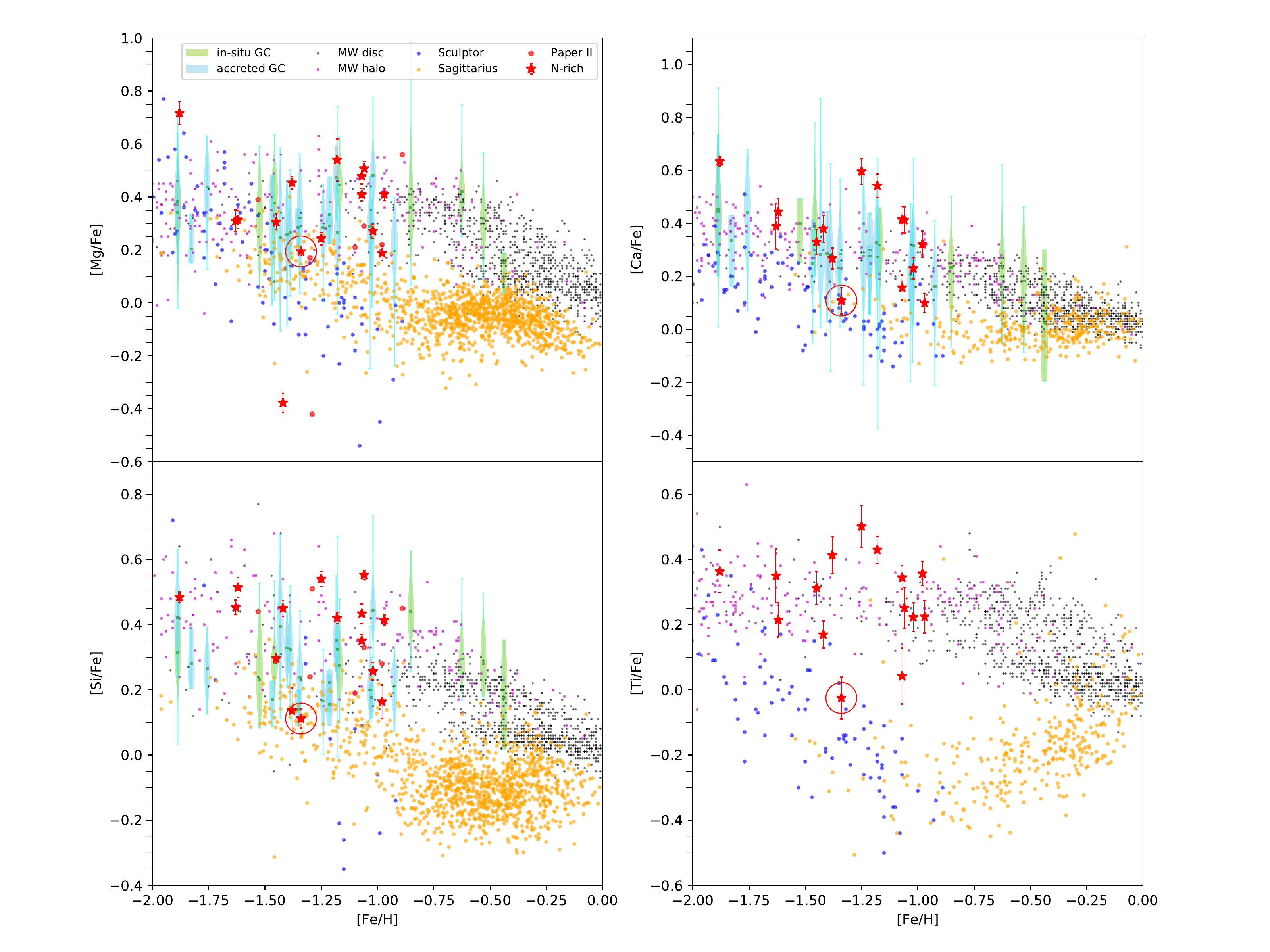}
	\caption{[Mg/Fe]-[Fe/H] (upper left), [Si/Fe]-[Fe/H] (bottom left), [Ca/Fe]-[Fe/H] (upper right), and [Ti/Fe]-[Fe/H] (bottom right) relation. Symbols are same as Figure~\ref{fig:Al_FeH}. Blue circles represent stars from Sculptor \citep{Hill2019}.}
	\label{fig:alpha_FeH}
\end{figure*}

The [Mg/Fe]-[Fe/H] relation is shown in the upper left panel of Figure~\ref{fig:alpha_FeH}. In the background, we also include MW halo and disc stars, dwarf galaxy (Sculptor and Sagittarius) stars, and APOGEE GC stars (violin shaped symbols).\footnote{There are 31 GCs compiled by \citet{Meszaros2020} in total. In our Figure~\ref{fig:Al_FeH} and \ref{fig:alpha_FeH}, we do not show Omega cen due to its large metallicity spread. Four GCs, which have metallicities smaller than -2.0, are not shown, neither. Among the rest 26 GCs, there are eight in situ GCs colored with yellow green and 18 accreted GCs colored with sky blue.} Each violin shaped symbol corresponds to one GC. The center value and min-max values are determined by its GC members selected from the APOGEE survey \citep[e.g.][]{ Meszaros2020, FT2020}. [Mg/Fe] of our N-rich field stars are inside the min-max range of APOGEE GCs of similar metallicity, which does not contradict with the hypothesis that N-rich field stars come from GCs. However, in situ GCs (GCs that identified as main disc or main bulge) and accreted GCs are largely overlapped especially in our chosen metallicity range ($-1.8 < {\rm [Fe/H]} < -1.0$), which prevents us distinguishing their in situ or accreted GC origin. Meanwhile, our N-rich field stars are mostly covered by both data points of MW stars and dwarf galaxies, indicating a mixture of in situ stars and extragalactic stars. This is consistent with their kinematics in \citetalias{Tang2020}.

As the atomic number increases, Si, Ca, and Ti become more stable against H-burning. Si abundance is suggested to vary by the ``Si-leakage'' during Mg-Al cycle when the core temperature reaches as high as $T \sim 65 \times 10^6\,{\rm K}$ \citep{Carretta2009b}. Most GCs do not show significant variation in Ca \citep[e.g.,][]{Carretta2010b}, except for several massive GCs with possible iron spread \citep{Carretta2020}. No significant Ti variation among GC members is reported to the best of our knowledge.

Interestingly, \citet{Horta2020} reported that in situ GCs may have higher [Si/Fe] compared to accreted GCs for ${\rm [Fe/H]} > -1.5$. The [Si/Fe] difference between two groups of GCs become less distinguishable as metallicity decreases. The background MW stars, dwarf galaxy stars (Sculptor and Sagittarius) of Figure~\ref{fig:alpha_FeH} (bottom left panel) support similar statement of \citet{Horta2020} for field stars, though the threshold metallicity can be only vaguely located at $-1.5 < {\rm [Fe/H]} < -1.0$. Upper right panel and bottom right panel of Figure~\ref{fig:alpha_FeH} shows the Ca and Ti abundances distributions of our N-rich field stars, along with background MW stars, dwarf galaxy stars and GC star distribution (violin shaped symbols), respectively. \cite{Jonsson2020} warned against using APOGEE-derived Ti abundances, since the Ti I and Ti II abundances from APOGEE pipeline may have unknown defects or large scatters, so [Ti/Fe] for APOGEE GCs are not shown in this work.

Looking at Mg, Si, Ca, Ti abundances versus [Fe/H] simultaneously, interesting results emerge:
(1) The $\alpha$-element difference between in situ (MW) stars and accreted (dwarf galaxy) stars is most significant for Ti, compared to Mg, Si and Ca. It is promising to use [Ti/Fe] for future works to distinguish accreted stars.
(2) The [$\alpha$/Fe] from APOGEE survey (violin shaped symbols) cover most of the N-rich field stars, again supporting their GC origin.
(3) Seven metal-rich ($-1.25 \le {\rm [Fe/H]} \le -0.95$) N-rich field stars show abundances more consistent with MW stars, indicating a higher possibility of in situ origin.
(4) At the lower end of the [$\alpha$/Fe] distribution, star \#9 (labeled with large red circle in Figure~\ref{fig:alpha_FeH}) shows consistently low Mg, Si, Ca, and Ti abundances similar to stars in dwarf galaxies, although it shows similar [Al/Fe] as MW disc and halo stars (Figure~\ref{fig:Al_FeH}). However, Al can be enhanced in GCs without altering much of its $\alpha$-abundances. Is it possible that this star has an extragalactic GC origin? We will further discuss this with more chemical information below.

\subsection{Iron-peak elements}

\begin{figure*}
	\centering \includegraphics[width=1.0\linewidth,angle=0]{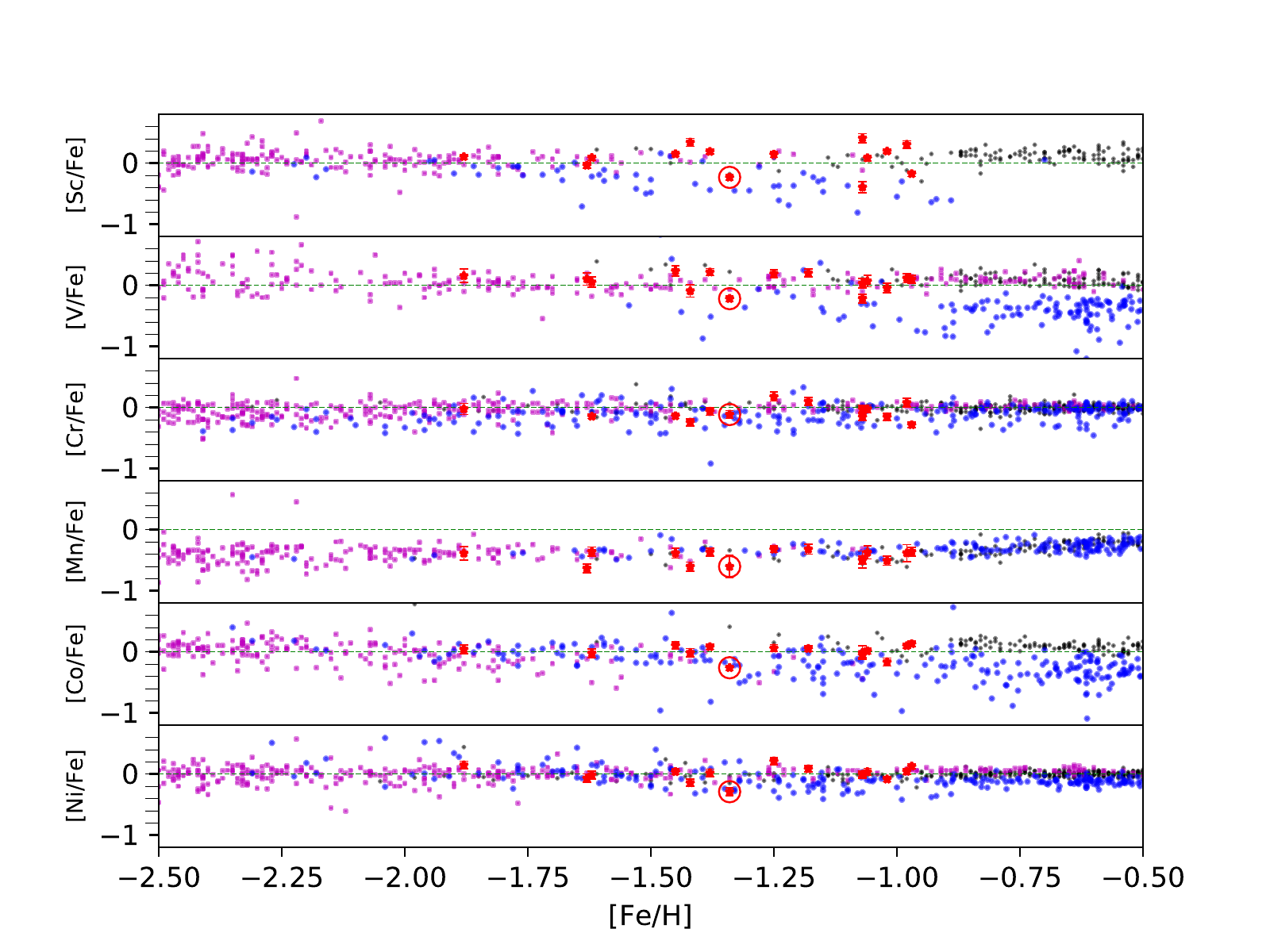}
	\caption{Abundances of iron peak elements vs. [Fe/H]. Abundances of N-rich stars are shown as red stars, while magenta squares and black dots correspond to MW stars from halo and disc, respectively (same sources as in Figure~\ref{fig:Al_FeH}). Blue circles correspond to stars from dwarf galaxies \citep[Sagittarius, Sculptor, Fornax, Carina, Leo I, ][]{Hasselquist2017, Hill2019, Shetrone2003}.}
	\label{fig:IronPeak_FeH}
\end{figure*}

Though Type Ia SNe, runaway deflagration obliterations of white dwarfs, have a signature more tilted towards the iron-peak group \citep{Nomoto1997}, the solar composition of the iron-peak elements are in fact a heterogeneous combination of both Type Ia SNe and core collapse Type II SNe \citep{Woosley1995}. As dwarf galaxies and MW have different star formation timescale, this discrepancy may manifest itself in iron-peak elements. Figure~\ref{fig:IronPeak_FeH} show that MW stars (black dots and magenta squares) and dwarf galaxy stars (blue circles) have appreciable different distributions in [Sc/Fe], [V/Fe], and [Co/Fe] vs. [Fe/H] graphs over the given metallicity range, especially in metal-rich part (${\rm [Fe/H]} \ge -1.25$). For these three elements, more metal-rich N-rich field stars ($-1.25 \le {\rm [Fe/H]} \le -0.95$) show distributions more similar to MW stars, confirming a higher possibility of in situ origin (Section~\ref{sec:alpha}). For [Cr/Fe], [Mn/Fe] and [Ni/Fe], MW stars and dwarf galaxy stars show less distinguishable difference. The discrepancies in iron-peak element of MW stars and dwarf galaxies are not evident in metal-poor part (especially for $-1.25 < {\rm [Fe/H]} < -1.5$) due to lack of data. However, if a linear relation of iron-peak abundances and metallicity is assumed for both MW (halo and disc) and dwarf galaxies, the dwarf galaxies would have lower iron-peak abundances. Therefore, star \#9 with low $\alpha$-element abundances also show lower abundances in [Sc/Fe], [V/Fe], [Co/Fe], agree with dwarf galaxies, which supports its possible extragalactic GC origin.

\subsection{Neutron capture elements}

Elements heavier than iron are generated via neutron-capture processes. Depending on the relative speed of neutron capture compared to $\beta$-decay, the neutron-capture processes are divided into rapid ({\it r}-) and slow ({\it s}-) ones. The main {\it s}-process elements are synthesized by AGB stars during thermal pulsations \citep{Busso2001, Karaks2014}. The astrophysical sites of producing {\it r}-process elements have been debated over the past sixty years \citep[e.g.,][]{Thielemann2011, Kajino2019}. The more popular models includes core collapse SNe \citep[e.g.,][]{Woosley1994} and neutron star mergers \citep[e.g.,][]{Cote2018, Watson2019}.

\begin{figure*}
	\centering \includegraphics[width=1.0\linewidth,angle=0]{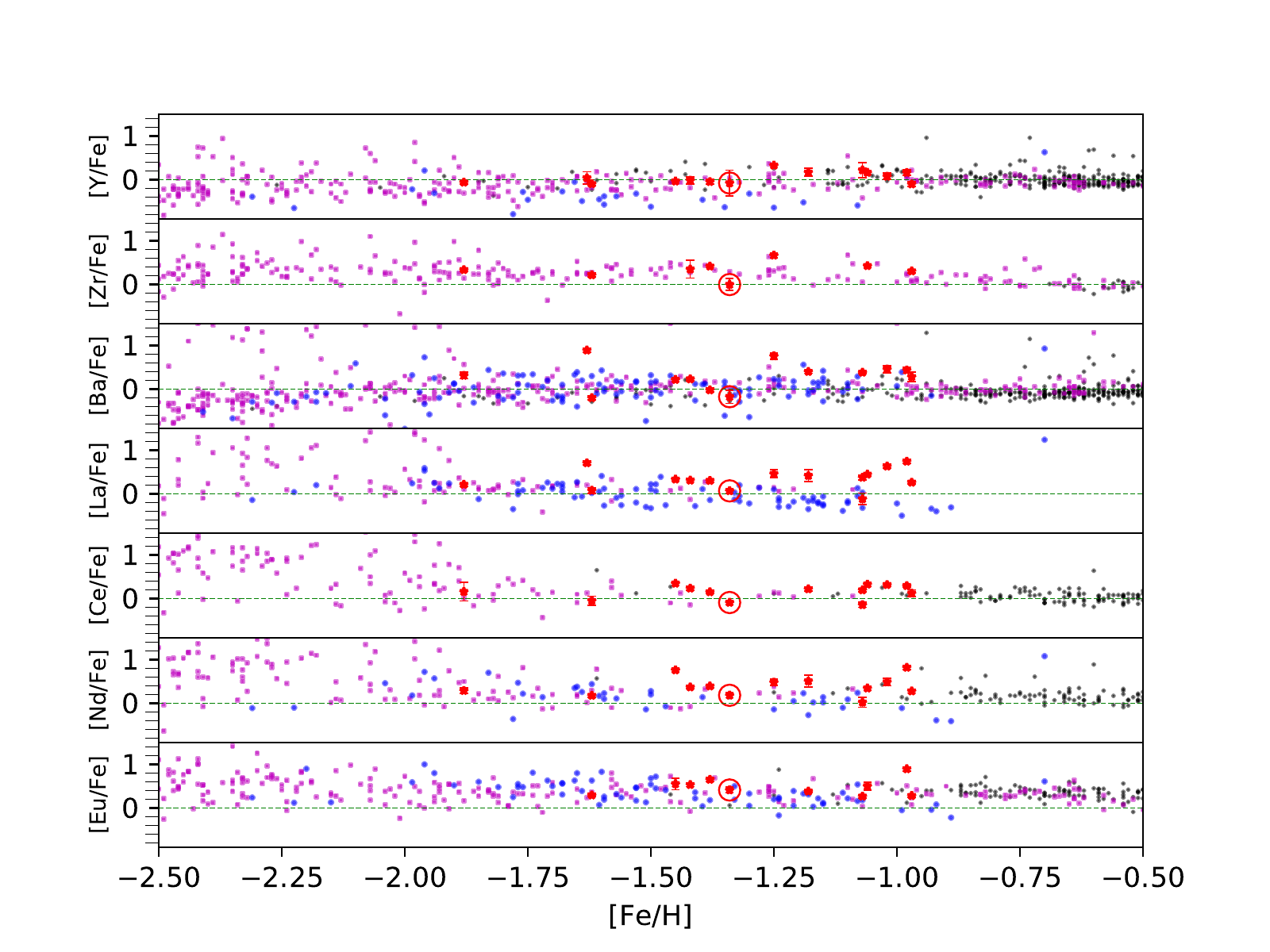}
	\caption{Abundances of neutron capture elements vs. [Fe/H]. Symbols are same as Figure~\ref{fig:IronPeak_FeH}.}
	\label{fig:Neutron_Capture}
\end{figure*}

Y, Zr, Ba, La, Ce, Nd and Eu are neutron-capture elements detectable in most of our sample stars across the observed wavelength. According to the recent work of \citet{Kobayashi2020}, Y, Zr, Ba, La, Ce and Nd are mostly produced through {\it s}-process in the current Universe, while Eu is produced through {\it r}-process. The neutron capture element abundances of our N-rich field stars are compared with MW stars in Figure~\ref{fig:Neutron_Capture}. We see that the scatter of abundances in MW stars increases as metallicity decreases, given that feature lines are weaker in more metal-poor stars. Generally, we see consistency between our N-rich field stars and MW stars, but several stars with enhanced [Ba/Fe], [La/Fe] and [Eu/Fe] are also noticed.

\begin{figure}
	\centering \includegraphics[width=1.0\linewidth,angle=0]{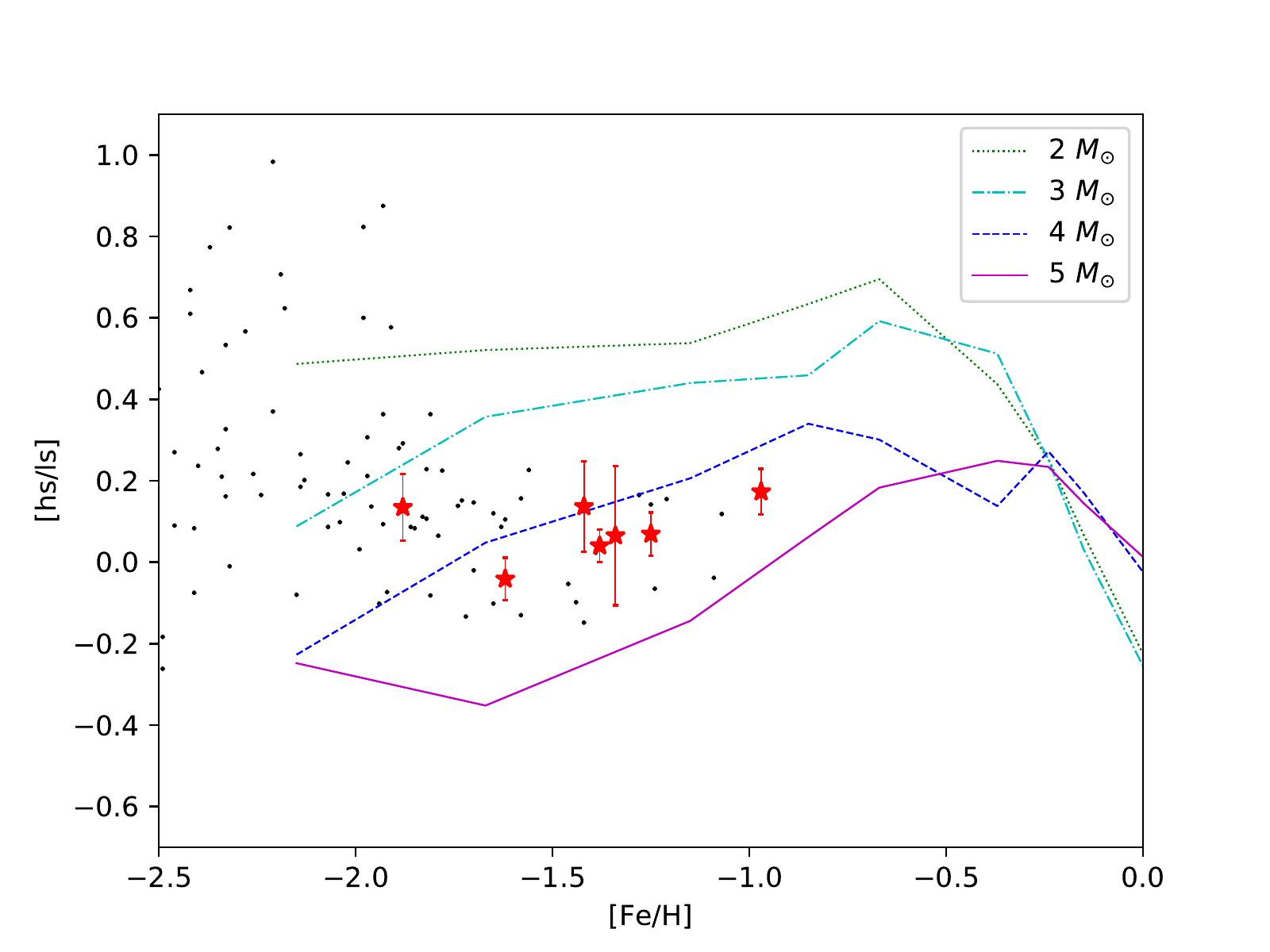}
	\caption{[hs/ls] vs. [Fe/H]. Abundances of N-rich stars are shown as red stars, while black dots correspond to Milky Way stars from halo and disc (same sources as black dots in Figure~\ref{fig:MgAl}). The grid lines with different colors and line styles indicate stars enriched by AGB stars of different masses \citep{Cristallo2015}.}
	\label{fig:hsls_FeH}
\end{figure}

Two nuclear reactions are the major neutron excess sources in AGB stars: $\rm {^{13}C(\alpha,n) ^{16}O}$ and $\rm {^{22}Ne(\alpha,n) ^{25}Mg}$. The first reaction dominates the low mass AGB stars, while the latter one is mainly found in massive AGB stars \citep{Cristallo2015}. As the number of free neutrons per iron seed increases, the {\it s}-process flow first seeds the light {\it s}-process peak (Sr$-$Y$-$Zr), extending to $\rm ^{136}Ba$, and then reaches the heavy {\it s}-process peak (Ba$-$La$-$Ce$-$Pr$-$Nd), extending to $\rm ^{204}Pb-^{207}Pb$ \citep{Bisterzo2014}. Therefore, the heavy {\it s}-process element to light {\it s}-process element ratio ([hs/ls]) is closely related to metallicity and initial stellar mass. In this work, [hs/ls] is defined as ${\rm [hs/Fe] - [ls/Fe]}$, where ${\rm [hs/Fe] = ([Ba/Fe]+[La/Fe]+[Nd/Fe])/3}$, and ${\rm [ls/Fe] = ([Y/Fe]+[Zr/Fe])/2}$. In our sample, there are six stars with all five {\it s}-process elements (Figure~\ref{fig:hsls_FeH}). Five stars show consistent [hs/ls] values as other MW stars in the same metallicity range. The grid lines of \citet{Cristallo2015} indicate these stars are enriched by AGB stars of masses around $3 - 5\, M_{\odot}$.

\begin{figure*}
	\centering \includegraphics[width=1.0\linewidth,angle=0]{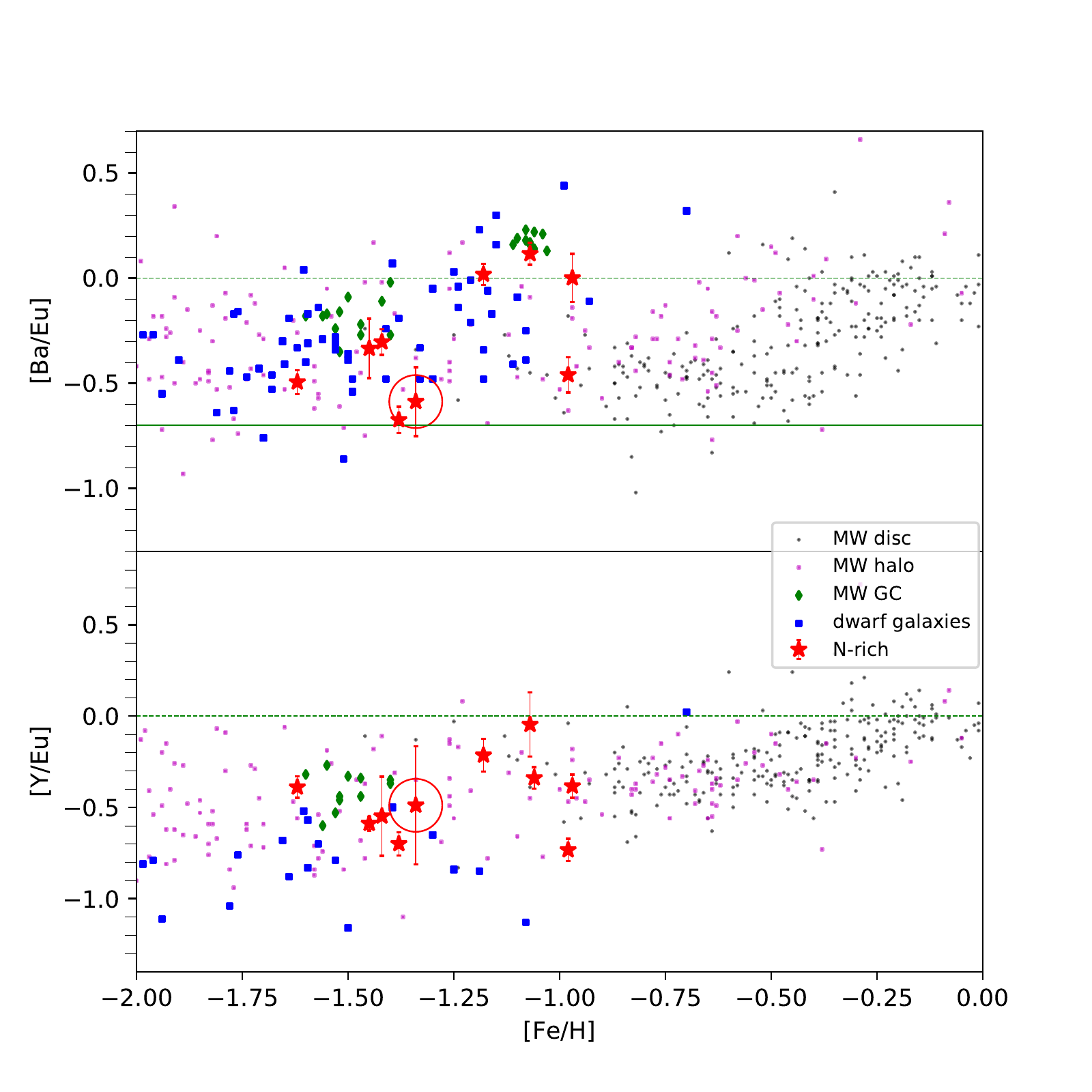}
	\caption{[Ba/Eu] vs. [Fe/H] and [Y/Eu] vs. [Fe/H]. Abundances of N-rich stars are shown as red stars.  Magenta squares correspond to MW halo stars, while black dots correspond to MW disc stars. The green solid line indicates that a typical value of a pure {\it r}-process by \citep{Bisterzo2014}.  Green diamonds correspond to MW GCs \citep{McWilliam1992, James2004, Munoz2013, Johnson2017, Massari2017}.  Blue squares correspond to dwarf galaxies from Sagittarius \citep{Hasselquist2017}, Sculptor \citep{Hill2019} (for [Ba/Eu]), Carina \citep{Venn2012}, and 19 red giants from Sagittarius, Fornax, Carina, Leo I \citep{Shetrone2003}.   Star \#9 with high extragalactic GC origin possibility are labeled with large red circle.}
	\label{fig:BaEu_YEu_FeH}
\end{figure*}

The ratio between {\it s}-process element and {\it r}-process element is an indicator of the neutron-capture speed compared to $\beta$-decay, which could be related to the contribution of AGB stars (thus SNe Ia) over SNe II and neutron star mergers. Ba is commonly used to represent the heavy {\it s}-process elements, while Y for the light {\it s}-process elements. A gradual rise in [Y/Eu] and [Ba/Eu] with increasing metallicity is seen in Figure~\ref{fig:BaEu_YEu_FeH}, which was addressed in e.g., \citet[][]{McWilliam1998}. The upper panel of Figure~\ref{fig:BaEu_YEu_FeH} shows that MW field stars, MW GC stars and dwarf galaxy stars overlap in the [Ba/Eu]-[Fe/H] space; In the lower panel, [Y/Eu] of dwarf galaxies from \citet{Shetrone2003} and \citet{Venn2012} are slightly smaller compared to MW field stars and MW GC stars at similar metallicity range, but with substantial overlap (also see Figure~19 of \citealt{Venn2012}). Though most N-rich field stars show consistent abundances with other studies, we are not able to distinguish their in situ or extragalactic origin here. \citet{Bisterzo2014} suggested that ${\rm [Ba/Eu]} \sim -0.7$ is the typical value for a pure {\it r}-process. Most N-rich field stars show [Ba/Eu] abundances higher than the pure {\it r}-process value (green solid line in Figure~\ref{fig:BaEu_YEu_FeH}), indicating a heterogeneous mixture of r- and s- processes. Star \#9 and star \#47 show [Ba/Eu] abundances close to the pure {\it r}-process value, indicating a strong r-process contribution.

\section{Discussion}
\label{sec:discussion}

\subsection{Chemical similarity between N-rich field stars and GC stars}
\label{subsec:sg}

Though our sample stars have been confirmed to be mostly N-rich using CN-CH features around 4000~\AA~\citepalias{Tang2019}, further chemical tagging with more elements requires high-resolution spectra. Using the MIKE and APOGEE spectra, we find that most N-rich field stars show consistent Na, O, Mg, Al, Si, and Ca with GC stars of similar metallicities. This chemical similarity supports the GC origin of these N-rich field stars. Two stars with strong Mg-depletion are also found. Strong Mg-depletion is usually found in very metal-poor ([Fe/H$]<-2.0$) GCs (e.g., M15, M92). Two N-rich field stars show strong Mg-depletion at [Fe/H$] \sim -1.3$ is somewhat puzzling. One possible scenario is: accreted materials from companion stars may change the Mg abundances of these two N-rich field stars dramatically \citep{Fernandez_Trincado2017}.

Based on the Na abundances, we find that most N-rich field stars are located in the transition region between FG and SG stars, thus it is difficult to verify their SG origin (see Figure~\ref{fig:NaO}). On the other hand, the Mg-Al figure (Figure~\ref{fig:MgAl}) shows that half of our sample have [Al/Fe$]>0.5$, which is largely different than GC FG stars. Interestingly, \citet{FT2020} identified 29 mildly metal-poor ([Fe/H$]<-0.7$) field stars with [Al/Fe$]>0.5$. They suggested that these stars were ejected into the bulge and inner halo from GCs formed in situ and/or GCs formed in dissolved dwarf galaxy progenitors.

\subsection{In situ or ex situ?}

\begin{figure}
	\centering \includegraphics[width=1.0\linewidth,angle=0]{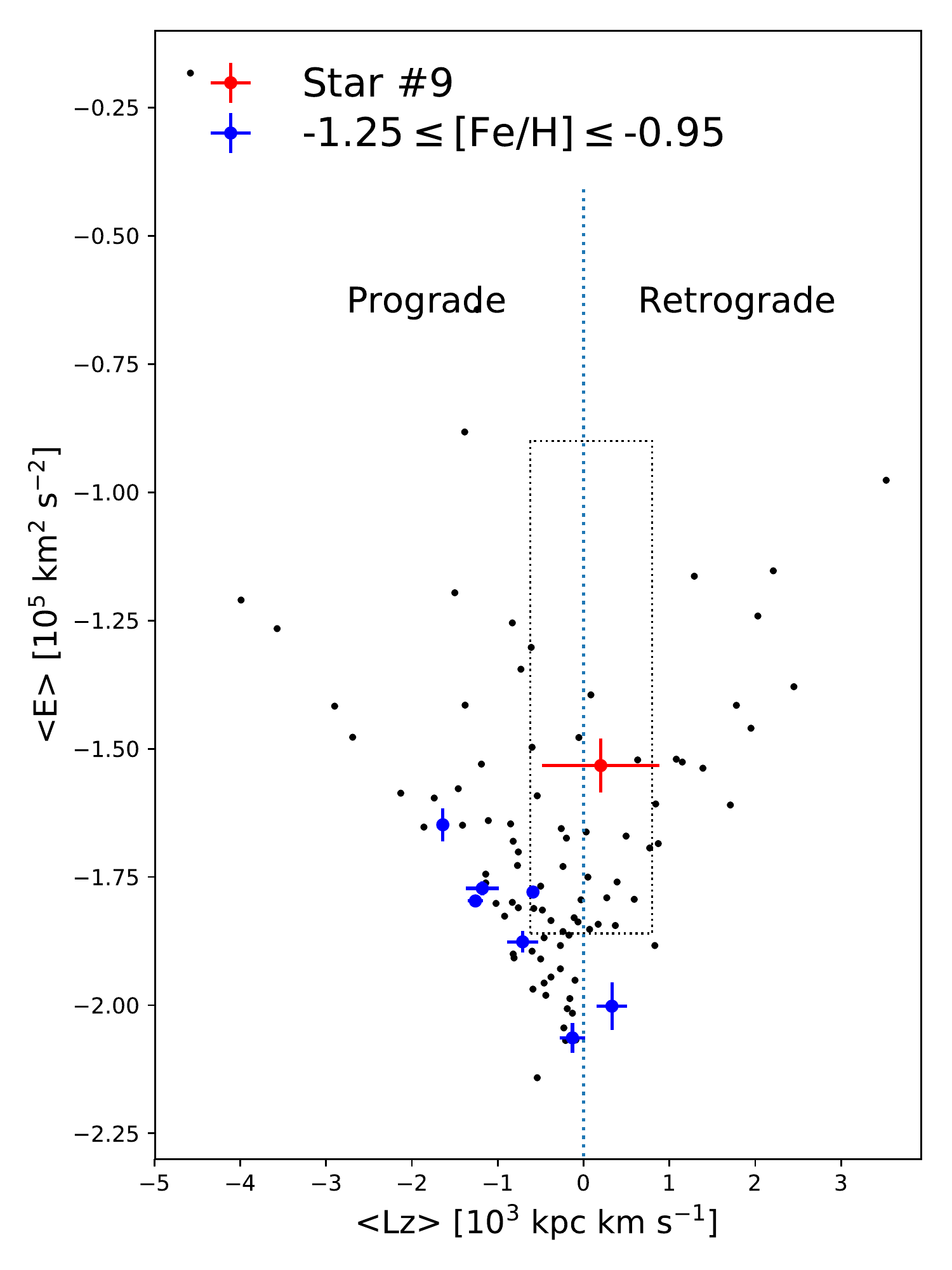}
	\caption{Mean $z$-direction angular momentum $<L_{\rm z}>$ vs. mean orbital energy $<E>$. Star \#9 is labelled as red dots with error bars, while stars with $-1.25 \le {\rm [Fe/H]} \le -0.95$ are labelled as blue dots with error bars. The dotted rectangle shows the GSE region$^a$ given by \citet{Massari2019}.}
	\label{fig:ELz}
	\begin{tablenotes}
	\item $^a$  Positive $<L_{\rm z}>$ indicate retrograde and negative $<L_{\rm z}>$ indicate prograde, which is the opposite in \citet{Massari2019}. This is related to the Galactocentric coordinate system that we used (\citetalias{Tang2020}).
	\end{tablenotes}
\end{figure}

As our understanding of the MW formation rapidly improves in the \emph{Gaia} era, GCs are now believed to formed in both MW and dwarf satellite galaxies, and later merged together as the current Galactic GCs \citep[e.g.,][]{Massari2019, Myeong2019}. GCs formed in situ and ex situ tend to show different chemical and dynamical signatures. If the N-rich field stars were escaped from GCs, they should carry similar chemo-dynamical information as the host GCs. \citetalias{Tang2020} showed that our parent N-rich field sample ($\sim 100$) have both stars formed in situ and stars formed ex situ using kinematics. In this work, one interesting star (star \#9, large circles in Figures~\ref{fig:alpha_FeH} and \ref{fig:IronPeak_FeH}) show consistently low [Mg/Fe], [Si/Fe], [Ca/Fe], [Ti/Fe], [Sc/Fe], [V/Fe], and [Co/Fe] compared to MW stars at similar metallicities, which is a strong evidence that this star was formed in dissolved dwarf galaxies. The relatively higher [Al/Fe] of star \#9 compared to dwarf galaxy field stars indicates that it is possibly enriched in GC environment, as Al-rich stars can be found in accreted GCs. On the other hand, stars with $-1.25 \le {\rm [Fe/H]} \le -0.95$ show [$\alpha$/Fe] and iron-peak abundances consistent with other MW field stars, indicating that they were formed in situ. Moreover, we checked their kinematic information in \citetalias{Tang2020}, and found that star \#9 shows $<E>$ and $<Lz>$ consistent with $Gaia$-Sausage-Enceladus (GSE) stars (Figure~\ref{fig:ELz}), while stars with $-1.25 \le {\rm [Fe/H]} \le -0.95$ show in situ disk or halo kinematics \citep[e.g.][]{Naidu2020}. Interestingly, star \#9 shows [Ba/Eu] close to the r-process limit, which is consistent with other GSE stars observed in \citet{Aguado2020}. This is a straightforward example to demonstrate that chemical tagging is possible to help decipher the origin of field stars.  We also acknowledge the difficulties in judging the origin of field stars based on several chemical and dynamical observation evidences, which are not always consistent with each other. A more sophisticated probability estimation of their origin is left for further studies.

\section{Conclusions}
\label{sec:conclusion}

We used MIKE spectra of 15 stars, which are selected from the LAMOST N-rich field star sample, to analyse their chemical abundances for more than 20 elements. We find that the Na, O, Mg, Al, Si, and Ca abundances of the N-rich field stars are consistent with GC stars, which supports their GC origin. Given that GC FG and SG stars have large overlap in Na-O and Mg-Al parameter space, it is difficult to confirm the SG origin of N-rich field stars. But we do find seven stars with ${\rm [Al/Fe]} > 0.5$, which is typical for GC SG stars.

We notice that one star (\#9) with consistently low [Mg/Fe], [Si/Fe], [Ca/Fe], [Ti/Fe], [Sc/Fe], [V/Fe], and [Co/Fe] show similar kinematic and [Ba/Eu] as other stars from GSE. On the other hand, more metal-rich stars ($-1.25 \le {\rm [Fe/H]} \le -0.95$) show $\alpha$-elements abundances and iron-peak abundances more consistent with MW field stars rather than dwarf galaxies, indicating likely in situ origin.

The ratio between heavy s-process elements and light s-process elements reveals that most N-rich field stars could be enriched by AGB stars with masses around $3 - 5\, M_{\odot}$. A detailed comparison between different chemical models and the obtained chemical patterns of our N-rich field stars would be fruitful for future discussion \citep[e.g.,][]{Masseron2020}.

\section*{Acknowledgments}
We thank Ian Thompson, Yang Huang for helpful discussions. We thank the anonymous referee for insightful comments.
J.Y. and B.T. gratefully acknowledges support from the National Natural Science Foundation of China under grant No. U1931102.
D.G. gratefully acknowledges support from the Chilean Centro de Excelencia en Astrof\'isica y Tecnolog\'ias Afines (CATA) BASAL grant AFB-170002.
D.G. also acknowledges financial support from the Direcci\'on de Investigaci\'on y Desarrollo de la Universidad de La Serena through the Programa de Incentivo a la Investigaci\'on de Acad\'emicos (PIA-DIDULS).
Guoshoujing Telescope (the Large Sky Area Multi-Object Fiber Spectroscopic Telescope LAMOST) is a National Major Scientific Project built by the Chinese Academy of Sciences. Funding for the project has been provided by the National Development and Reform Commission. LAMOST is operated and managed by the National Astronomical Observatories, Chinese Academy of Sciences.

\bibliography{nrich}
\bibliographystyle{aasjournal}

\appendix

\begin{figure}
	\centering \includegraphics[width=1.0\linewidth,angle=0]{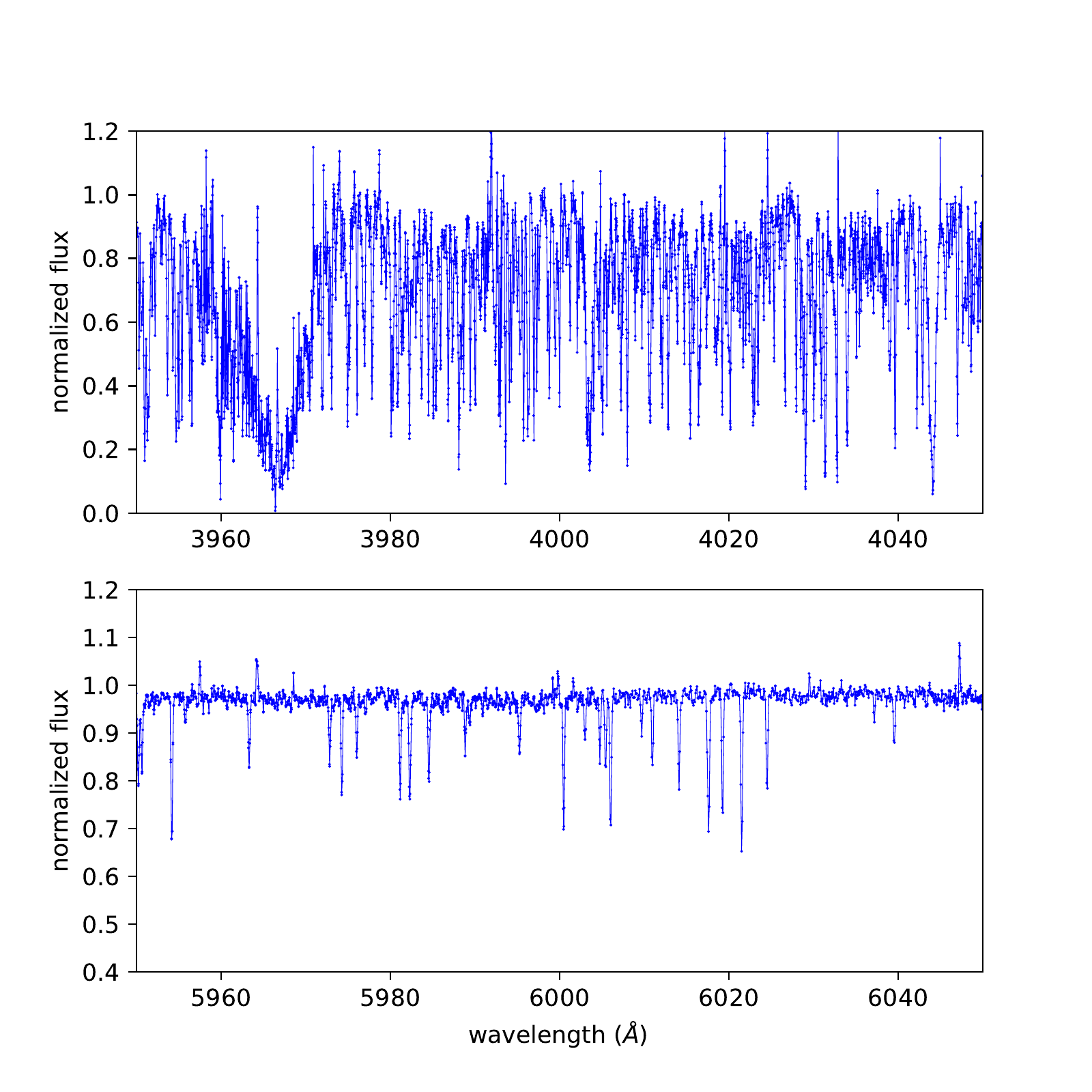}
	\caption{Spectra of star \#88 of both blue side and red side.}
	\label{fig:spectra}
\end{figure}

\begin{figure}
	\centering \includegraphics[width=1.0\linewidth,angle=0]{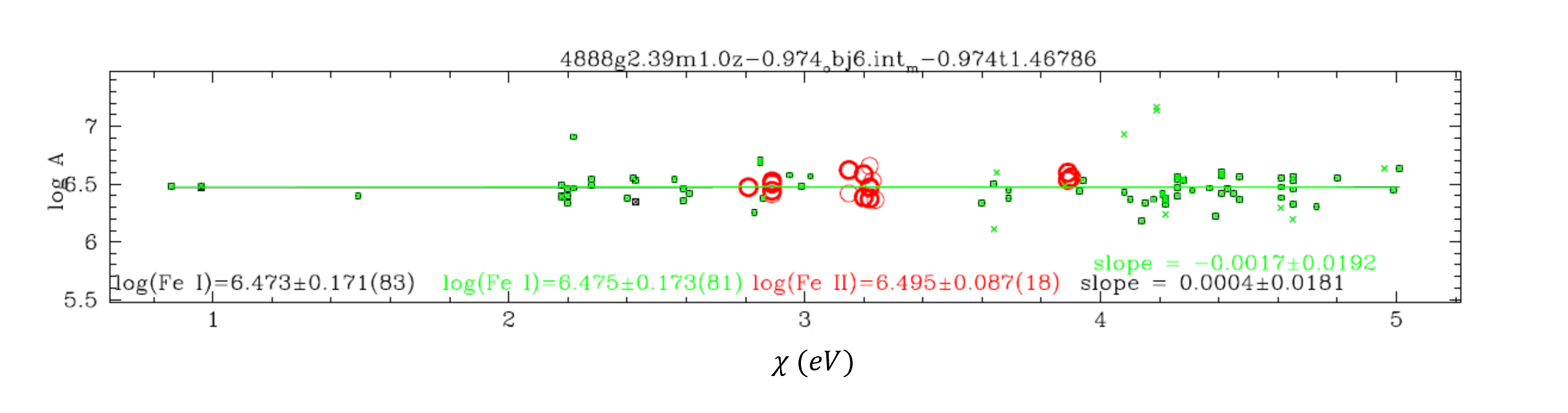}
	\caption{Illustration of $T_{\rm eff}$ determination with iron abundances and excitation potential.}
	\label{fig:Teff_determ}
\end{figure}

\begin{figure}
	\centering \includegraphics[width=1.0\linewidth,angle=0]{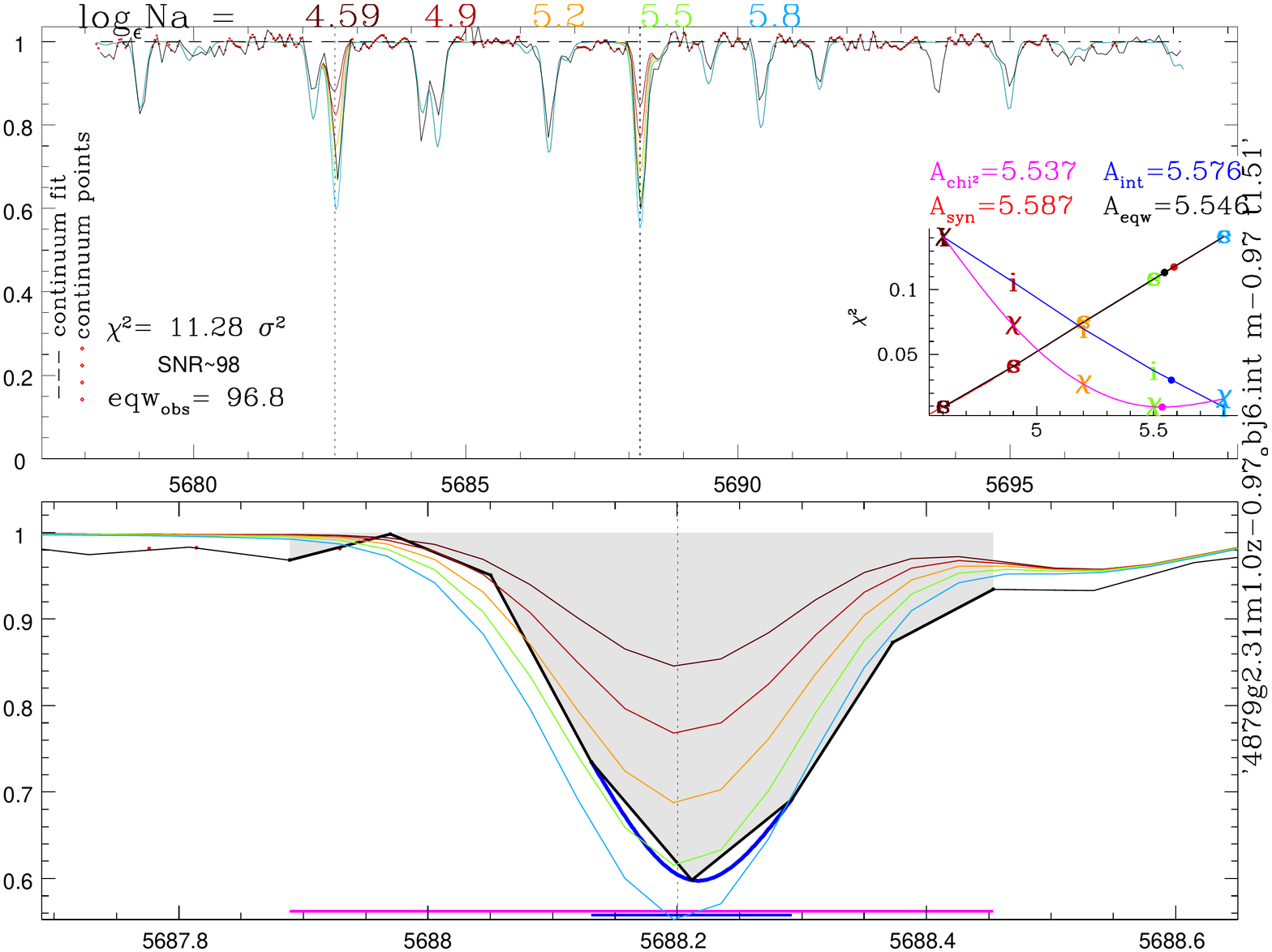}
	\caption{Comparison of four different methods ($\chi^2$ minimization, line intensity, equivalent width, and spectral synthesis) in determining Na abundances for star \#88. The black solid line shows the observed normalized spectrum, while colored lines show synthesis obtained by different methods in both upper and bottom panel. The inserted panel in the bottom right corner of the upper panel shows the diagram for the four methods for abundances determination. Each of the method is represented by a different color ($\chi^2$ minimization: magenta, line intensity: blue, equivalent width: black, spectral synthesis: red). (See more detailed descriptions in \citep{Masseron2016}.)}
	\label{fig:spec_abund}
\end{figure}

Spectra of star \#88 of both blue side ($3950 - 4050$ \AA) and red side ($5950 - 6050$ \AA) are shown in Figure~\ref{fig:spectra}. We adopt red side spectra ($\lambda>5000\,\text{\AA}$) in abundances determination when considering their better SNR. $T_{\rm eff}$ is determined by obtaining null trend for the abundance of iron lines against excitation potential is shown in Figure~\ref{fig:Teff_determ}. An example of abundances determination with four different methods is shown in Figure~\ref{fig:spec_abund} as mentioned in Section~\ref{sec:sample}.

\end{document}